\documentclass[conference]{IEEEtran}
\IEEEoverridecommandlockouts

\usepackage{cite}
\usepackage{amsmath,amssymb,amsfonts}
\usepackage{graphicx}
\usepackage{textcomp}
\usepackage{xcolor}
\usepackage{booktabs}
\usepackage{multirow}
\usepackage{array}
\graphicspath{{figures/}}

\usepackage[colorlinks=true,urlcolor=blue,linkcolor=black,citecolor=black]{hyperref}
\usepackage{eso-pic}

\begin{document}

\title{Sub-6 GHz Over-the-Air AMC via Curriculum Fine-Tuned CNN-Transformers}

\author{
\IEEEauthorblockN{Nurettin Safak, Muhammet Sefa Demirel, Alperen Marasli, Taha Eren Atmaca, Durdu Can Yerdeyatar, Ozgun Ersoy}
\IEEEauthorblockA{Department of Electrical and Electronics Engineering, Ankara Yildirim Beyazit University, Ankara, Turkey}
\IEEEauthorblockA{nsafaked@gmail.com, sefademirel56@gmail.com, alperen.eem@gmail.com,\\
tahaeren1883@gmail.com, yrdeytr4601@gmail.com, ozgun.ersoy@aybu.edu.tr}
}

% Copyright/IEEE pubid: \IEEEpubid'in otomatik boşluk ayırma mekanizması bu
% belgede metinle çakışıyordu. Bunun yerine sayfa 1'e MUTLAK konumlu (metin
% akışından bağımsız) bir kutu yerleştiriliyor -- hiçbir gövde metniyle
% asla çakışmaz (bildiri_rf'te kullanılan çözümle aynı).
\AddToShipoutPictureFG*{%
  \ifnum\value{page}=1
    \AtPageLowerLeft{%
      \hspace{0.7in}\raisebox{0.55in}{%
        \footnotesize\textbf{979-8-3195-2149-1/26/\$31.00~\textcopyright{}2026 IEEE}%
      }%
    }%
  \fi
}

\maketitle

\begin{abstract}
Automatic modulation classification (AMC) models are frequently trained and validated on synthetic or channel-cabled data, leaving open the question of how they behave once path loss and antenna pointing error are introduced by a genuine free-space link. We report a curriculum fine-tuning study of a hybrid CNN-Transformer AMC model. The general-purpose, all-32-class dataset underlying the model was built entirely at 915\,MHz, on a controlled, clock/PPS-synchronized MIMO-expansion-cable link (not spatial-multiplexing transmission); all subsequent free-space, real-hardware experimentation -- the sequential fine-tuning curriculum, matched-distance evaluation, and every reported over-the-air accuracy figure -- was carried out at 4\,GHz, across five directional-antenna distances (25, 35, 50, 70, 75\,cm) under fixed TX/RX gain. Three distances used near-ideal antenna alignment ($\approx$99\%) and two used a deliberately introduced partial misalignment ($\approx$85\%), fine-tuned last in the curriculum. We report matched-distance test accuracy (91.8--93.7\% across all five 4\,GHz conditions) and confusion-matrix analysis grounded in RF theory, and report honestly where our sequential fine-tuning order confounds cumulative link adaptation with antenna alignment, rather than overstating what the data can support.
\end{abstract}


% ════════════════════════════════════════════════════════════════
\section{Introduction}

Spectrum monitoring and electronic-warfare (EW) systems increasingly rely on deep-learning-based automatic modulation classification (AMC) to identify signals in real time. The large majority of reported AMC results, however, are obtained on synthetic datasets or on cabled/simulated channels that do not expose the model to free-space path loss, hardware gain limitations, or the antenna pointing error that a real directional link -- particularly one on a moving platform -- inevitably introduces. Foundational work by O'Shea~\emph{et al.}~\cite{oshea2018ota} first demonstrated that deep AMC models trained in simulation degrade under real over-the-air impairments (carrier offset, multipath, fading), and more recent hybrid architectures validated on real SDR captures~\cite{padhya2025cnnlstm} confirm the gap persists. A model that performs well on clean, synthetic I/Q data offers no guarantee that it will remain reliable once these real-world degradations are present, and few studies report a controlled transition from a clean reference link to a genuine over-the-air (OTA) deployment at GHz-range carrier frequencies, with antenna pointing error further degrading directional EW links~\cite{dabiri2022pointing,zhao2025uavjamming}.

This work reports such a transition. Prior work has explored an uncertainty-driven hybrid architecture (2D-CNN with Bayesian MLP and a BiLSTM fallback stage) for broad-coverage RF modulation recognition~\cite{safak2026uncertainty}; the present study instead adopts a different hybrid CNN-Transformer design, better suited to the sequential curriculum fine-tuning problem studied here. We draw a clear line between the two roles frequency plays in this study: 915\,MHz is used exclusively to build the general, all-32-class base dataset and train the model under controlled, cable-synchronized conditions, while 4\,GHz is where all real-world, free-space experimentation takes place -- the hybrid CNN-Transformer AMC architecture, once trained at 915\,MHz, is adapted through a sequential fine-tuning curriculum to a real 4\,GHz free-space link across five distances, two of which include deliberately introduced antenna misalignment representative of platform tracking error. Our goal is to determine whether a model whose fundamental modulation characteristics were learned on a clean reference link can be efficiently adapted -- by updating only its temporal-modeling stage while keeping learned low-level RF feature extraction fixed -- to a physically distinct, degraded, real-world link, and to report matched-distance performance and failure modes honestly, including where our own experimental design does not permit clean causal attribution.

% ════════════════════════════════════════════════════════════════
\section{System Model}

\subsection{Motivation for a Hybrid CNN-Transformer Design}

Modulation classification from raw I/Q samples requires two qualitatively different kinds of structure: local amplitude/phase transitions between adjacent samples (informative for constellation-based digital modulations) and longer-range temporal dependencies that span a large fraction of the observation window (informative for classes such as frequency-hopping spread spectrum, FHSS, whose defining structure is a hopping pattern rather than any single short segment). A convolutional stack is well matched to the first kind of structure but has a limited effective receptive field; a Transformer encoder is well matched to the second but, applied directly to a 1024-sample raw I/Q sequence, is computationally costly and lacks an inductive bias for local signal structure. We therefore use a hybrid design: a convolutional stem for cheap, local feature extraction, downsampling the sequence before it reaches a Transformer encoder that models the longer-range dependencies over the reduced sequence. Related hybrid convolutional-Transformer designs have recently been proposed for AMC, combining CNN feature extraction with attention-based or graph-based long-range modeling~\cite{wang2023ctgnet,ruikar2024hctc}; our design follows the same general hybrid principle but is evaluated end-to-end on a real, physically degraded OTA link rather than on simulation alone.

\subsection{Multi-Task Hybrid CNN-Transformer Architecture (``Plan A'')}

\begin{figure}[!t]
\centering
\includegraphics[width=0.99\columnwidth]{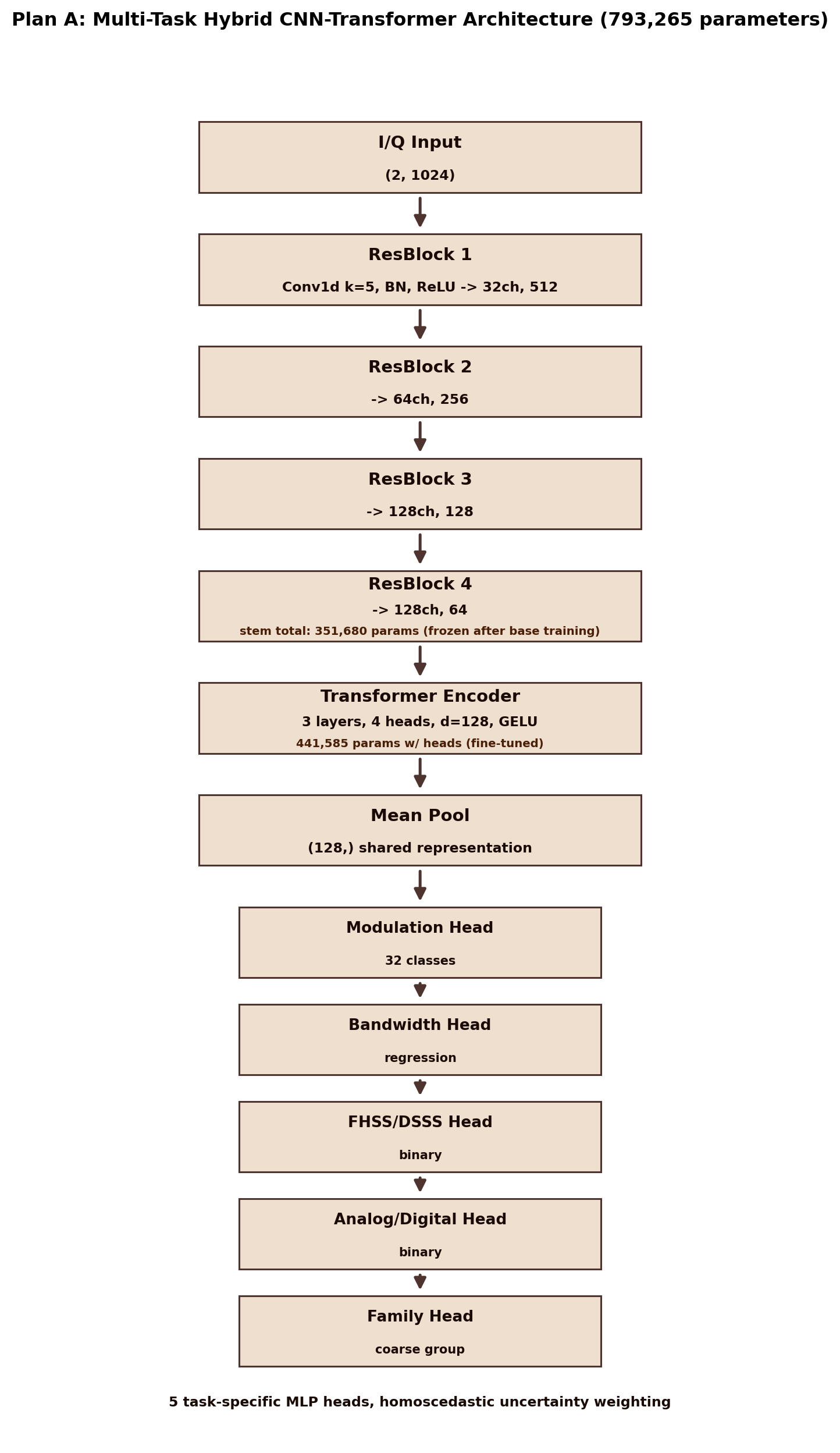}
\caption{Plan A architecture: I/Q input, four-block 1D-ResNet stem, Transformer encoder, mean pooling, and five task-specific heads. 793,265 parameters total.}
\label{fig:arch}
\end{figure}

Figure~\ref{fig:arch} shows the architecture. The input is a raw I/Q tensor of shape $(2, 1024)$. A 1D-ResNet stem of four residual blocks (Conv1d, kernel size 5, batch normalization, ReLU, each block halving the sequence length while increasing channel width up to $d_{\mathrm{model}}=128$) reduces the sequence from 1024 to 64 time steps, extracting local amplitude/phase transition features (351,680 parameters). A sinusoidal positional encoding is added, followed by a Transformer encoder (3 layers, 4 attention heads, $d_{\mathrm{model}}=128$, feedforward dimension 256, GELU activation, dropout 0.1) that models the long-range temporal dependencies over the 64-step sequence. The pooled (mean over time) 128-dimensional representation feeds five task-specific heads (each a two-layer MLP, 441,585 parameters combined with the Transformer): (i) modulation classification (32 classes), (ii) bandwidth regression, (iii) FHSS/DSSS spread-spectrum detection, (iv) analog/digital discrimination, and (v) coarse modulation-family classification. The family head was added after observing, in earlier live-hardware testing, that the model's confusions were overwhelmingly between adjacent orders of the same family (e.g., 8/16/32-PSK, 32/64/128-APSK, 64/128/256-QAM, OOK/4/8-ASK); forcing the shared trunk to also predict family membership indirectly regularizes it toward family-discriminative features, aiding the fine-grained head.

The five task losses are combined via homoscedastic uncertainty weighting~\cite{kendall2018}: rather than hand-tuning fixed loss weights, each task $i$ has a learnable log-variance $s_i$, and the total loss is
\begin{equation}
\mathcal{L} = \sum_{i=1}^{5} \left( e^{-s_i} \mathcal{L}_i + s_i \right),
\end{equation}
which lets the optimizer automatically down-weight noisier or harder-to-fit tasks over training rather than requiring a manual weighting schedule. More recent analysis has examined the limitations of this homoscedastic weighting scheme and proposed analytical alternatives~\cite{kirchdorfer2025uncertainty}; we retain the original formulation here for its simplicity and adopt its limitations as a candidate direction for future refinement.

% ════════════════════════════════════════════════════════════════
\section{Experimental Setup}

\begin{figure}[!t]
\centering
\includegraphics[width=0.99\columnwidth]{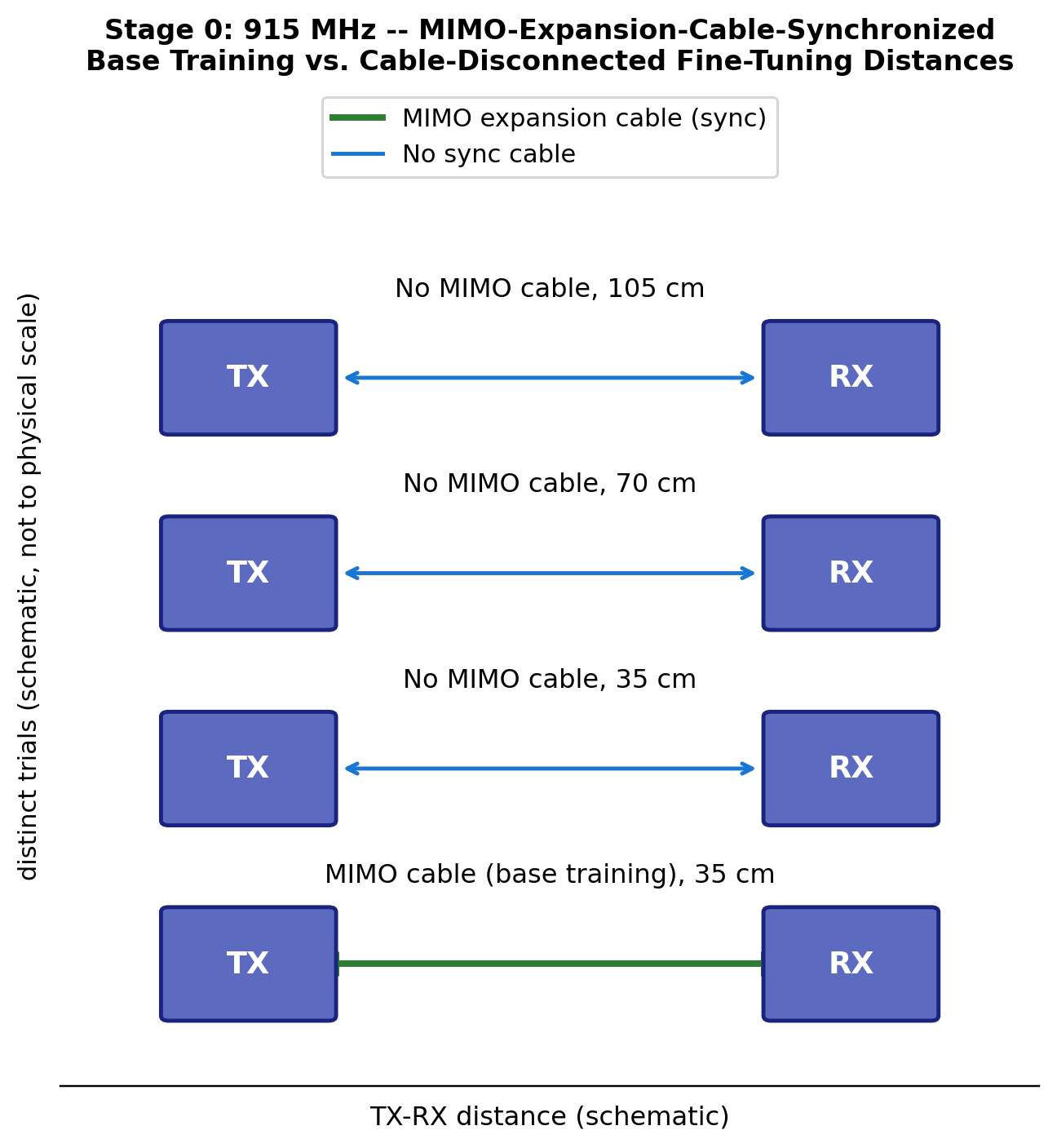}
\caption{Experimental setup, Stage 0: 915\,MHz base training used a MIMO expansion cable for clock/PPS synchronization between the two radios (not multi-antenna MIMO transmission); subsequent cable-disconnected distances informed the earlier stage of this research program.}
\label{fig:setupA}
\end{figure}

\begin{figure}[!t]
\centering
\includegraphics[width=0.99\columnwidth]{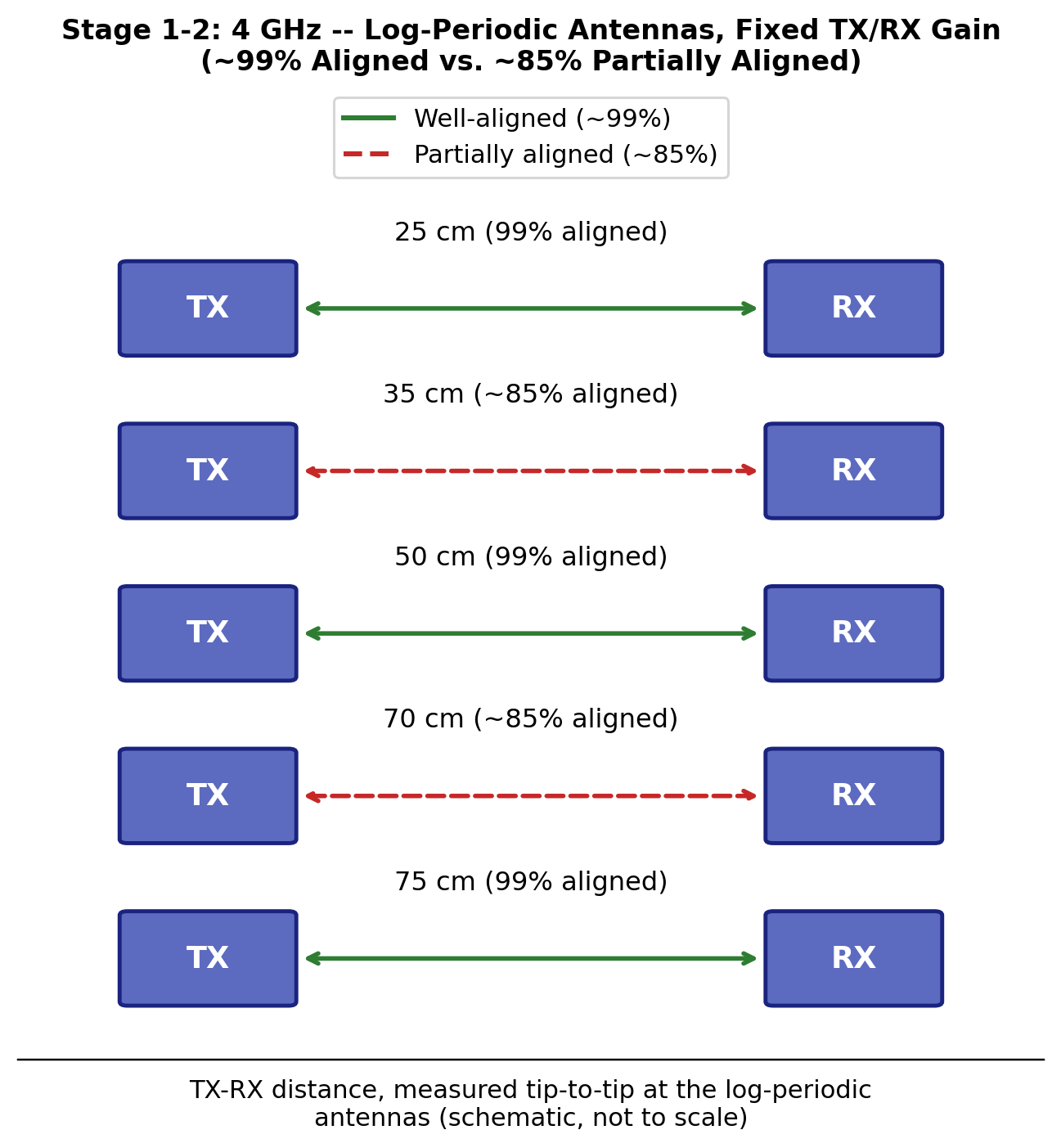}
\caption{Experimental setup, Stages 1--2: 4\,GHz fine-tuning used log-periodic directional antennas at five distances, three well-aligned (25, 50, 75\,cm) and two partially aligned (35, 70\,cm), distances measured tip-to-tip.}
\label{fig:setupB}
\end{figure}

Figures~\ref{fig:setupA} and~\ref{fig:setupB} show the physical test bench for both stages: two SDRs (USRP N-series) facing each other, connected either via a MIMO expansion cable (Stage 0) or purely over the air (Stages 1--2).

\begin{figure*}[!t]
\centering
\includegraphics[width=0.99\textwidth]{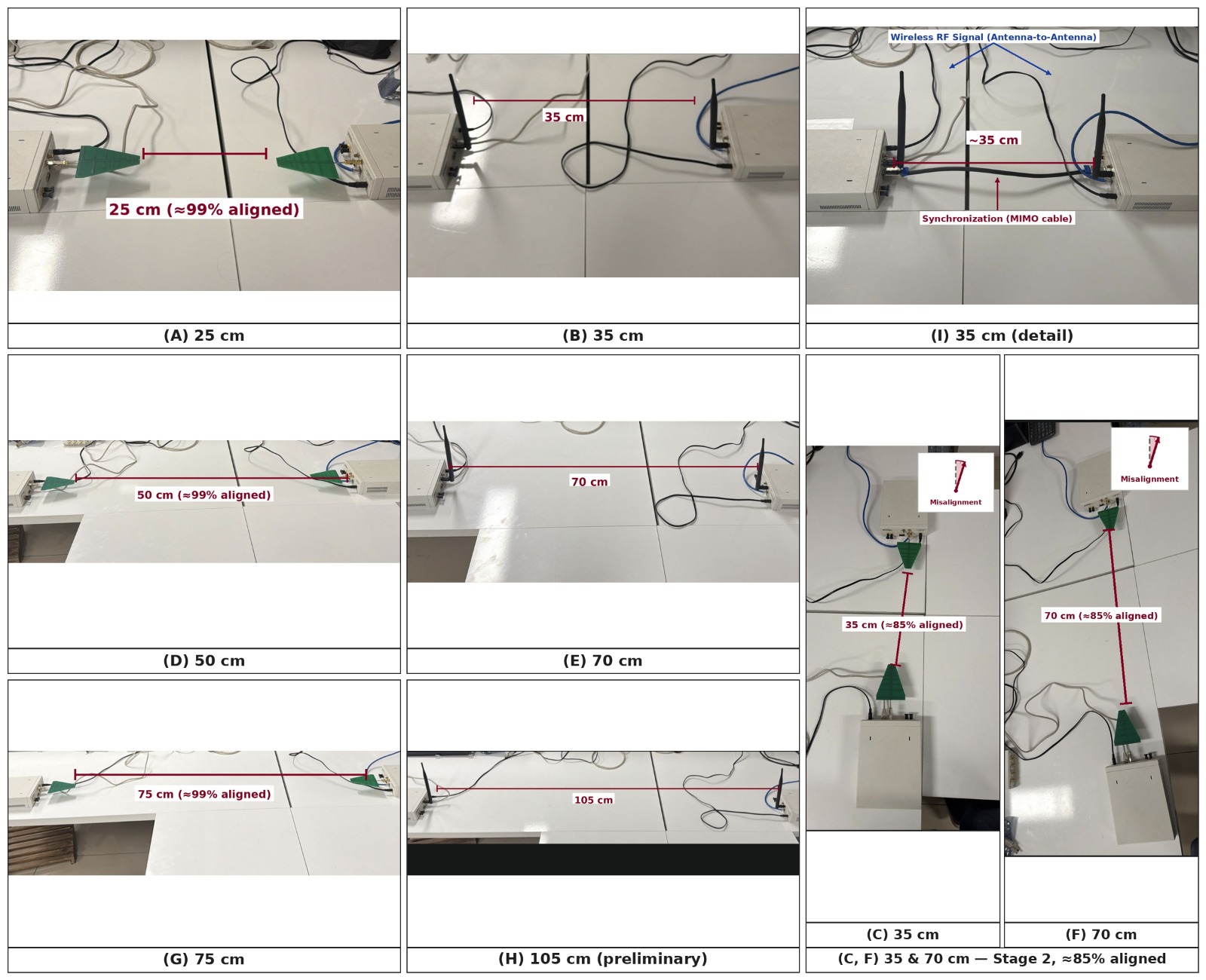}
\caption{Physical realization of the setups described in Section~III. (A,D,G) Stage~1 distances (25, 50, 75\,cm), near-ideal boresight alignment ($\approx$99\%). (C,F) Stage~2 distances (35, 70\,cm), deliberately introduced partial misalignment ($\approx$85\%); the inset protractor glyph is a schematic indicator of the qualitative misalignment condition, not a measured angle. (B,E) Alternate top-down views of the same two Stage~2 links. (H) The 105\,cm preliminary range-finding link discussed in Section~III-B, which was excluded from the formal evaluation sweep. (I) Detail view of the Stage~0 configuration, illustrating that the MIMO expansion cable (bottom) provides only clock/PPS synchronization while the RF signal itself travels over the air between the two antennas (top), as described in Section~III-A.}
\label{fig:setupPhotos}
\end{figure*}

\subsection{Stage 0: 915\,MHz Cable-Synchronized Base Training}

The base dataset was collected at 915\,MHz with the two USRP N-series radios connected via a MIMO expansion cable, which shares a reference clock and PPS timing signal between the TX and RX units for coherent synchronization -- it does not indicate multi-antenna spatial-multiplexing transmission; the link is single-TX/single-RX (SISO) throughout. This clean, synchronized configuration serves as the controlled ``laboratory'' condition in which the model first learns the fundamental characteristics of all 32 modulation classes across its five tasks. To generate variable-SNR training labels, TX gain was swept over a fixed ladder (0--10\,dB across SNR labels) while RX gain was held at its default (5\,dB); 32 modulation classes, $\approx$1.85M training windows. The base model was trained for 30 epochs; a separate 100-epoch run confirmed this budget is sufficient, with validation accuracy peaking at epoch 34 (94.59\%) and no further improvement observed through epoch 95 before the run was stopped.

\subsection{Stages 1--2: 4\,GHz Free-Space Adaptation}

The MIMO expansion cable was disconnected and the link moved to 4\,GHz free-space transmission through directional log-periodic UWB antennas, introducing free-space path loss and (at two of five distances) antenna pointing error that the 915\,MHz cable-synchronized stage did not contain. Because this link is gain-limited at 4\,GHz, TX gain was fixed at 31.5\,dB (hardware ceiling) and RX gain at 15\,dB for \emph{all} distances and SNR labels, so that distance and alignment -- rather than a TX-power-based SNR ladder as in Stage~0 -- are the varying factors.

Preliminary range-finding tests extended up to 105\,cm to observe the absolute limits of the physical link. However, to maintain strict control over alignment variables and ensure reliable performance attribution, the formal evaluation sweep was bounded at a maximum of 75\,cm. Thus, data were collected at five primary distances: 25, 35, 50, 70, and 75\,cm. At 25, 50, and 75\,cm (Stage~1) the directional antennas were aligned as close to ideal boresight as physically achievable ($\approx$99\%). At 35 and 70\,cm (Stage~2), the antennas were \emph{deliberately} left at a partial alignment of $\approx$85\% of ideal, to emulate the pointing/tracking error that a moving platform (e.g., a maneuvering UAV or a mobile jammer) would realistically exhibit under an EW engagement.

% ════════════════════════════════════════════════════════════════
\section{Training Strategy: Sequential Curriculum Fine-Tuning}

\begin{figure}[!t]
\centering
\includegraphics[width=0.99\columnwidth]{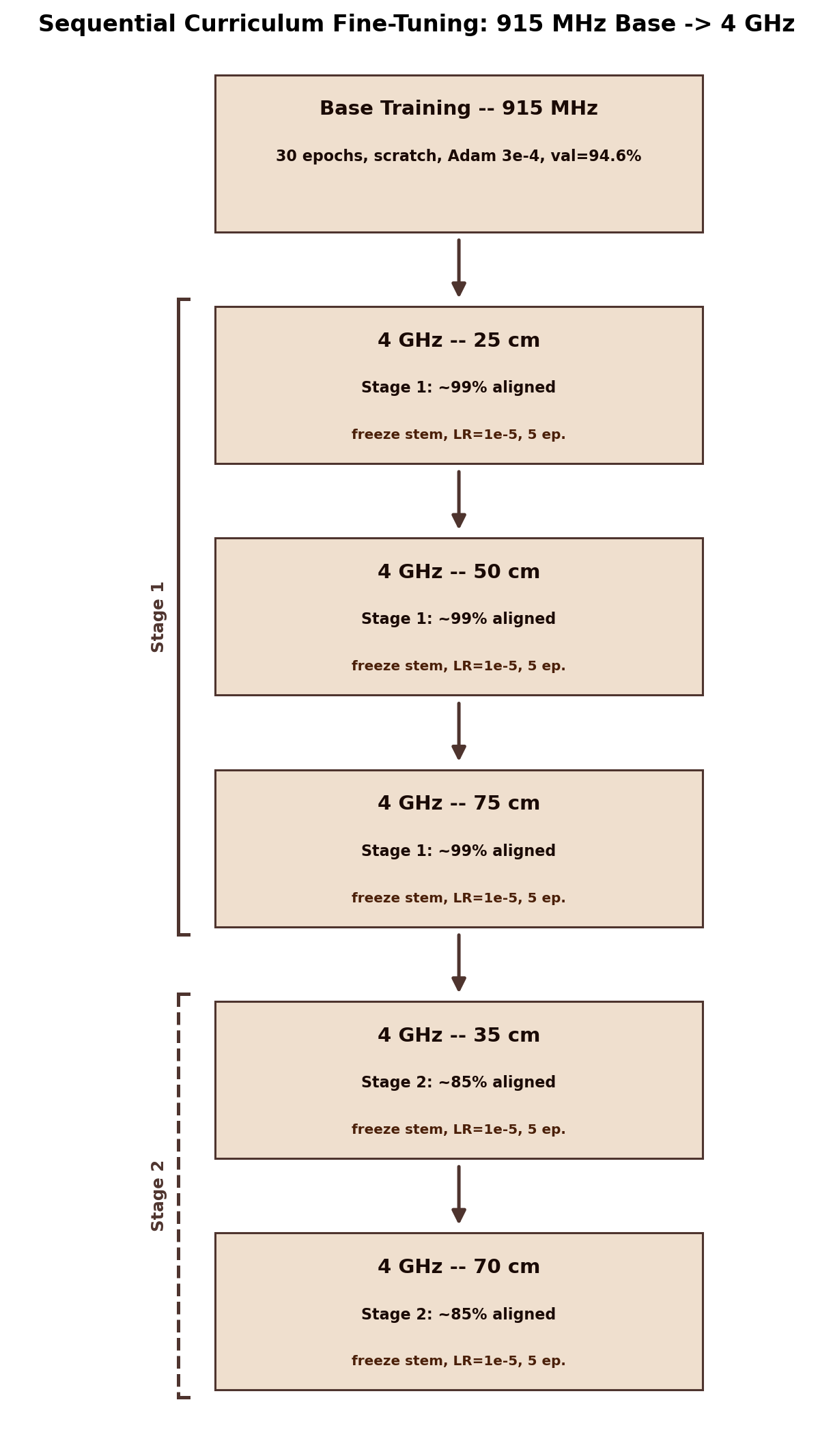}
\caption{Sequential curriculum fine-tuning: the Stage~0 (915\,MHz) checkpoint is fine-tuned through Stage~1 (well-aligned 4\,GHz distances) and then Stage~2 (partially-aligned distances), each step freezing the CNN stem and updating only the Transformer and task heads.}
\label{fig:flow}
\end{figure}

Rather than treating each 4\,GHz distance as an isolated fine-tuning target, we adopt a sequential curriculum (Fig.~\ref{fig:flow}) in the spirit of curriculum learning~\cite{bengio2009curriculum}, ordering conditions from well-aligned to partially-misaligned rather than presenting them in random or arbitrary order: starting from the Stage~0 checkpoint, the model is fine-tuned first through the three well-aligned distances (25\,cm~$\rightarrow$~50\,cm~$\rightarrow$~75\,cm, Stage~1), then through the two partially-aligned distances (35\,cm~$\rightarrow$~70\,cm, Stage~2), each stage's checkpoint initializing the next. This progressive fine-tuning across changing hardware and channel conditions follows the broader logic of transfer learning between radios and link conditions explored in~\cite{muller2024transfer,wong2024rftl}, applied here specifically to a cabled-to-free-space frequency transition. At every step, the 1D-ResNet stem -- which encodes low-level RF amplitude/phase transition features learned from the large, clean 915\,MHz dataset -- is frozen, and only the Transformer encoder and the five task heads are updated (5 epochs, AdamW, $\eta=10^{-5}$, cosine annealing), with the checkpoint retained only if held-out validation accuracy improves over its immediate predecessor. This reflects a specific hypothesis: that the low-level feature representation learned once, on abundant clean data, need not be re-learned for a new physical channel, and that only the temporal/contextual reasoning stage requires adaptation to a new link's characteristics. Each resulting checkpoint is evaluated on its own distance's held-out test split (matched-distance evaluation), reporting accuracy, macro-F1, and the full 32-class confusion matrix.

We note explicitly that this curriculum design means distance, alignment condition, and cumulative fine-tuning exposure are not independently controlled variables in this study.

% ════════════════════════════════════════════════════════════════
\section{Results and Discussion}

\subsection{Matched-Distance Accuracy}

Table~\ref{tab:distance_results} summarizes matched-distance test accuracy (192,000 test windows per condition) after fine-tuning at each of the five 4\,GHz distances.

\begin{table}[!h]
\caption{Matched-Distance Test Accuracy (4\,GHz, Fixed TX/RX Gain)}
\label{tab:distance_results}
\centering
\small
\begin{tabular}{lccc}
\toprule
\textbf{Distance} & \textbf{Alignment} & \textbf{Accuracy} & \textbf{Macro-F1} \\
\midrule
25\,cm (Stage 1) & $\approx$99\% & 92.23\% & 0.921 \\
50\,cm (Stage 1) & $\approx$99\% & 92.86\% & 0.928 \\
75\,cm (Stage 1) & $\approx$99\% & 91.81\% & 0.916 \\
\midrule
35\,cm (Stage 2) & $\approx$85\% & 93.67\% & 0.935 \\
70\,cm (Stage 2) & $\approx$85\% & 93.11\% & 0.930 \\
\bottomrule
\end{tabular}
\end{table}

Across all five conditions, matched-distance accuracy stays within a narrow 91.8--93.7\% band, indicating that the curriculum fine-tuning strategy keeps the model well-adapted to the 4\,GHz free-space link throughout the distance sweep and into the partially-aligned Stage~2 conditions, without a collapse in performance. We report this range as the primary finding rather than a specific ranking among the five points: within Stage~1 (well-aligned), accuracy does not decrease monotonically with distance (92.23\%~$\rightarrow$~92.86\%~$\rightarrow$~91.81\% at 25, 50, 75\,cm), and the Stage~2 (partially-aligned) points score numerically \emph{higher} than every Stage~1 point (93.67\% at 35\,cm, 93.11\% at 70\,cm).

We deliberately do not interpret the Stage~2 numbers as evidence that partial misalignment improves, or fails to harm, accuracy. Because Stage~2 checkpoints are fine-tuned \emph{after}, and initialized from, the Stage~1 checkpoints, they have received strictly more cumulative fine-tuning exposure to the 4\,GHz link than any Stage~1 checkpoint. Our curriculum design therefore confounds antenna alignment with cumulative fine-tuning order, and this experiment cannot isolate how much of the observed Stage~2 performance reflects robustness to partial misalignment specifically versus the benefit of additional adaptation steps on the same link. What we can report, without overreaching, is that the model maintains 91.8--93.7\% matched-distance accuracy across the full curriculum, including its final, partially-misaligned conditions -- a result consistent with, but not proof of, robustness to the alignment errors introduced at Stage~2.

\subsection{Confusion Analysis}

\begin{figure}[!t]
\centering
\includegraphics[width=0.99\columnwidth]{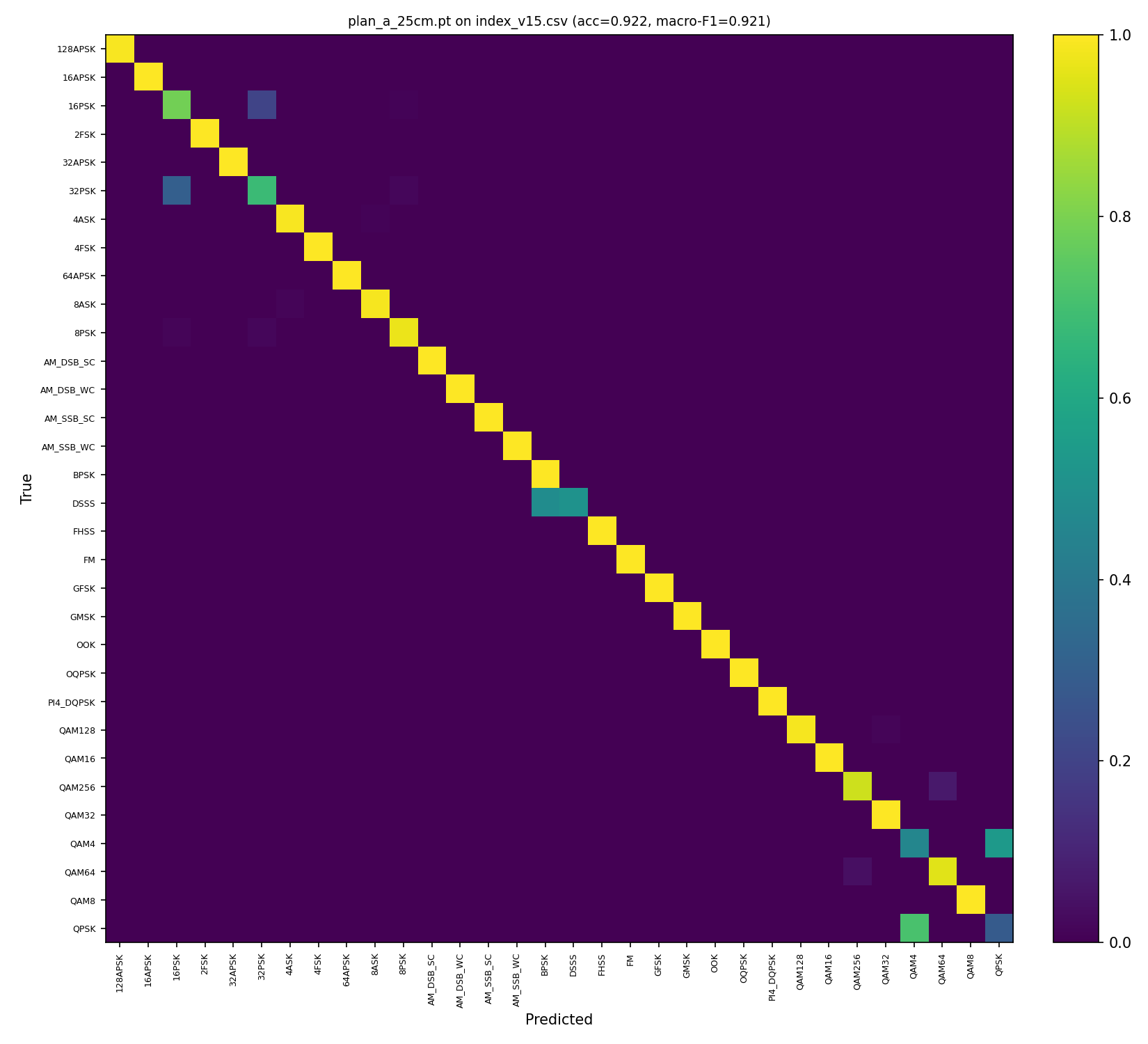}
\caption{Confusion matrix, best case: 25\,cm, $\approx$99\% aligned.}
\label{fig:cm25}
\end{figure}

\begin{figure}[!t]
\centering
\includegraphics[width=0.99\columnwidth]{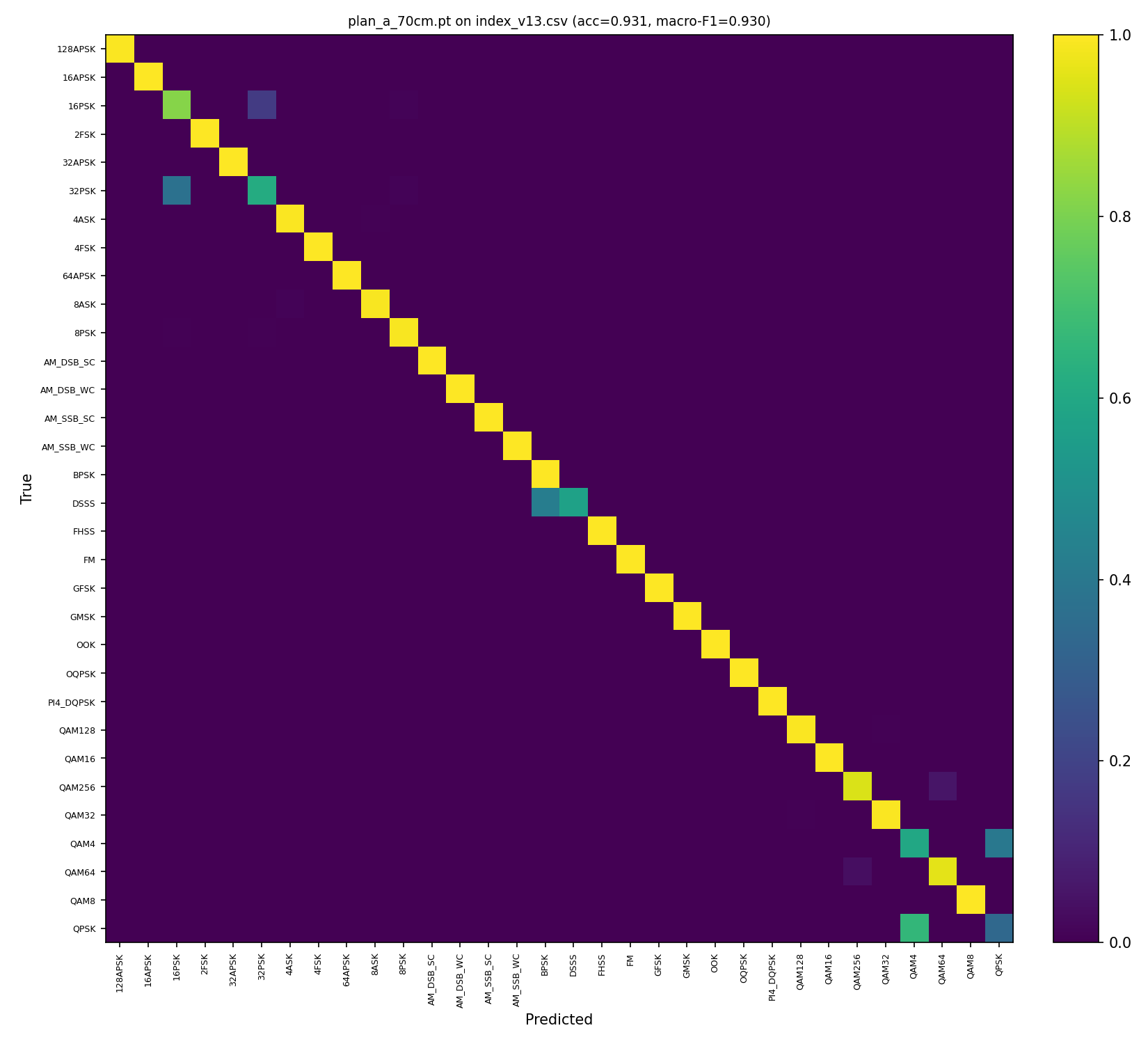}
\caption{Confusion matrix, challenging case: 70\,cm, $\approx$85\% aligned.}
\label{fig:cm70}
\end{figure}

Figures~\ref{fig:cm25} and~\ref{fig:cm70} show the confusion matrices for the best-case (25\,cm, $\approx$99\% aligned) and one challenging (70\,cm, $\approx$85\% aligned) condition. The dominant, consistent source of error across all five conditions is the same: QPSK is confused with QAM4 (4-QAM) far more often than it is correctly classified (e.g., at 25\,cm, 4282 of 6000 QPSK test windows were predicted as QAM4, versus 1714 correct). This is not a model failure specific to distance or alignment -- QPSK and 4-QAM correspond to the same four-point constellation under different modulation-family naming conventions, so this confusion is physically expected from the class taxonomy itself. Similarly, DSSS is frequently confused with BPSK (2922 of 6000 at 25\,cm) because the DSSS signals in this dataset are BPSK-chip-modulated spread spectrum; absent explicit despreading, the two are difficult to distinguish from short-window I/Q statistics alone. Both confusions appear at comparable rates across all five distance/alignment conditions, indicating they are structural properties of the class taxonomy and the 1024-sample window length rather than channel-condition-dependent effects.

\begin{figure}[!t]
\centering
\includegraphics[width=0.99\columnwidth]{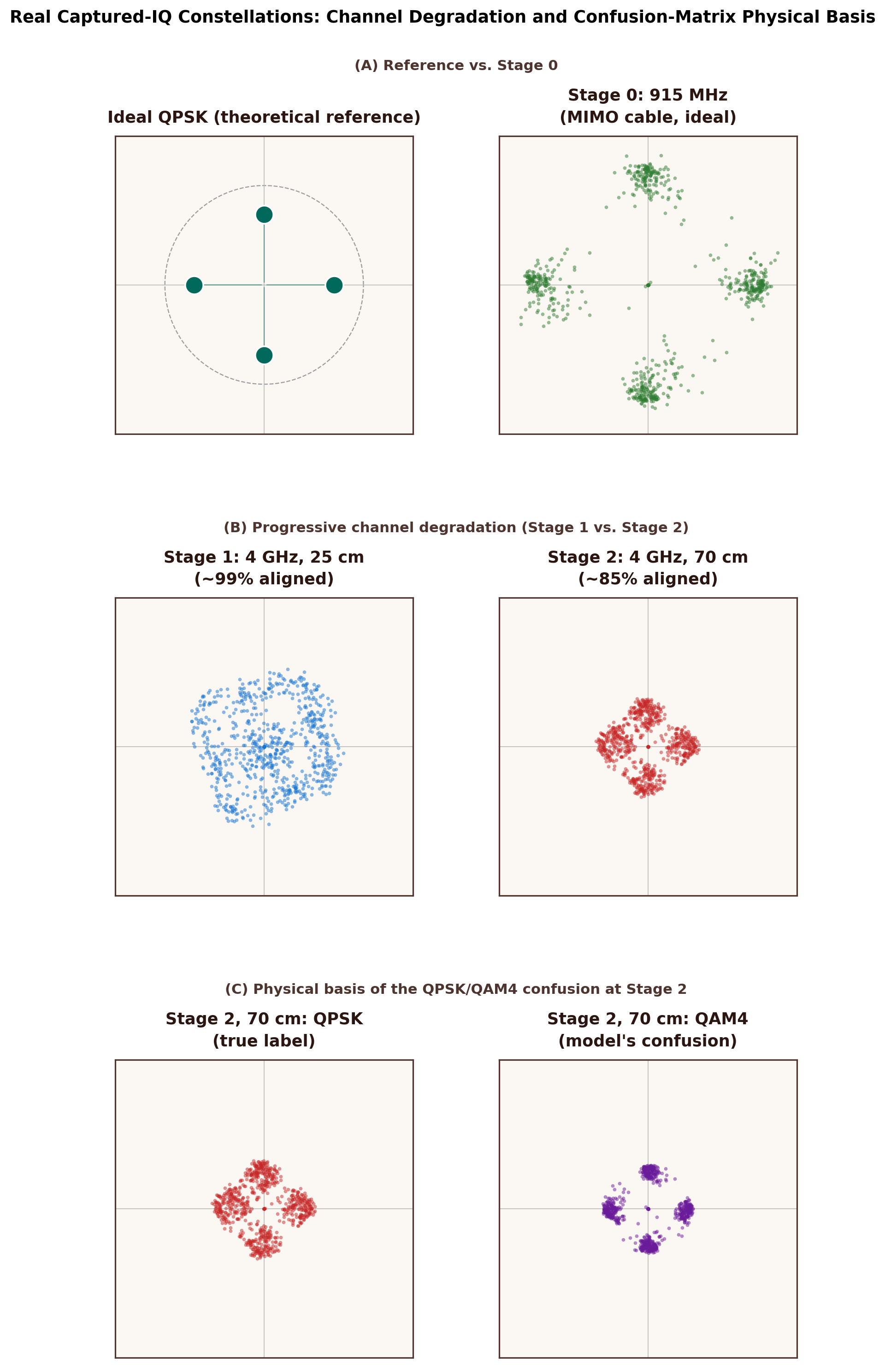}
\caption{Real captured-I/Q constellations grounding the QPSK/QAM4 confusion in the physical channel. (A) Theoretical ideal QPSK vs.\ Stage~0 (915\,MHz, cable-synchronized): tight, well-separated clusters. (B) Progressive degradation of the same modulation (QPSK) from Stage~1 (25\,cm, $\approx$99\% aligned) to Stage~2 (70\,cm, $\approx$85\% aligned) under free-space path loss and alignment-induced phase scattering. (C) At 70\,cm, the QPSK (true label) and QAM4 (model's confusion) constellations physically overlap in the I/Q plane -- the confusion is a consequence of channel degradation and class taxonomy, not a classifier failure.}
\label{fig:constellation_grid}
\end{figure}

Figure~\ref{fig:constellation_grid} grounds this confusion directly in the physical I/Q domain, using real captured I/Q throughout except where explicitly marked theoretical. Panel~(A) contrasts a theoretical ideal QPSK constellation with real captured I/Q from Stage~0 (915\,MHz, MIMO-cable-synchronized): the four clusters are tight and well-separated, indicating that under a controlled, low-noise channel the model's underlying feature representation has the capacity to resolve the class correctly. Panel~(B) traces the same modulation (QPSK) as the link moves to 4\,GHz free-space at 25\,cm ($\approx$99\% aligned) and then 70\,cm ($\approx$85\% aligned): the clusters broaden and rotate as free-space path loss lowers the received SNR and antenna pointing error introduces phase scattering, degrading the once-tight constellation into overlapping angular smears. Panel~(C) states the causal argument directly: at 70\,cm, the QPSK constellation (true label) and the QAM4 constellation the model confuses it with occupy the same region of the I/Q plane. The corresponding entries in the confusion matrix are therefore not attributable to an algorithmic deficiency in the classifier; they reflect a hardware-alignment-induced degradation that pushes two classes already sharing an identical four-point taxonomy (Section~V-B) into physical overlap, a condition under which no classifier operating on the I/Q signal alone can be expected to separate them.

\subsection{Ablation Studies}

To justify three architecture and training design choices used in Plan~A, we conducted three ablations against the same Stage~0 base-training protocol and, where applicable, the same five-distance curriculum.

\textbf{Transformer contribution (ResNet-only baseline).} We trained a Transformer-free variant of Plan~A -- identical 1D-ResNet stem, task heads, and multi-task loss, with the Transformer encoder removed -- on the Stage~0 dataset for 40 epochs. This variant reached a best validation modulation accuracy of 92.32\%, versus 94.60\% for the full hybrid model under the same base-training protocol (Table~\ref{tab:ablation_resnet}). This 2.28-point gap is consistent with our motivation for including a Transformer encoder (Section~II-A): capturing long-range temporal dependencies across the modulation window provides a measurable accuracy benefit beyond what the convolutional stem alone extracts.

\begin{table}[!h]
\caption{Transformer Contribution: Hybrid vs.\ ResNet-only (Stage 0)}
\label{tab:ablation_resnet}
\centering
\small
\begin{tabular}{lcc}
\toprule
\textbf{Variant} & \textbf{Epochs} & \textbf{Val.\ Mod.\ Acc.} \\
\midrule
Hybrid CNN-Transformer (Plan A) & 30 & 94.60\% \\
1D-ResNet-only (no Transformer) & 40 & 92.32\% \\
\bottomrule
\end{tabular}
\end{table}

\textbf{Multi-task vs.\ single-task supervision.} We repeated the full five-distance curriculum starting from a single-task variant of the Stage~0 checkpoint (modulation loss only, no auxiliary bandwidth/CFO/analog-digital/family heads). Table~\ref{tab:ablation_multitask} shows multi-task supervision outperforming single-task supervision at four of the five distances, by an average of 0.55 points overall (92.74\% vs.\ 92.19\%); the single exception is 75\,cm, where single-task scores marginally higher (92.20\% vs.\ 91.81\%). This supports retaining the auxiliary tasks as a regularizing signal on the shared representation rather than as incidental outputs, while acknowledging the effect is not perfectly uniform across conditions.

\begin{table}[!h]
\caption{Multi-Task vs.\ Single-Task Supervision (Matched-Distance Accuracy)}
\label{tab:ablation_multitask}
\centering
\small
\begin{tabular}{lcc}
\toprule
\textbf{Distance} & \textbf{Multi-task (Plan A)} & \textbf{Single-task} \\
\midrule
25\,cm & 92.23\% & 91.49\% \\
50\,cm & 92.86\% & 91.96\% \\
75\,cm & 91.81\% & 92.20\% \\
35\,cm & 93.67\% & 92.40\% \\
70\,cm & 93.11\% & 92.92\% \\
\midrule
\textbf{Average} & \textbf{92.74\%} & \textbf{92.19\%} \\
\bottomrule
\end{tabular}
\end{table}

\textbf{Freeze-stem trade-off.} We repeated the curriculum a third time without freezing the 1D-ResNet stem, updating the full network (793{,}265 parameters) rather than only the Transformer and task heads (441{,}585 parameters) at each fine-tuning step. Contrary to our stated hypothesis (Section~IV) that the low-level RF feature representation would not need re-learning, the no-freeze variant scored marginally higher on average (93.09\% vs.\ 92.74\%, Table~\ref{tab:ablation_freeze}), and on four of the five distances. We report this honestly as a partial refutation of the freezing hypothesis rather than omit it: freezing the stem does not improve accuracy in our setting. We nonetheless retain stem-freezing in Plan~A, since its practical justification is a $\sim$44\% reduction in updated parameters per fine-tuning step -- relevant for a system intended to adapt on constrained onboard hardware -- rather than an accuracy gain, and the accuracy cost of doing so is small ($<$0.4 points on average).

\begin{table}[!h]
\caption{Freeze-Stem vs.\ No-Freeze (Matched-Distance Accuracy)}
\label{tab:ablation_freeze}
\centering
\small
\begin{tabular}{lcc}
\toprule
\textbf{Distance} & \textbf{Freeze-stem (Plan A)} & \textbf{No-freeze} \\
\midrule
25\,cm & 92.23\% & 92.84\% \\
50\,cm & 92.86\% & 92.95\% \\
75\,cm & 91.81\% & 93.08\% \\
35\,cm & 93.67\% & 92.45\% \\
70\,cm & 93.11\% & 94.15\% \\
\midrule
\textbf{Average} & \textbf{92.74\%} & \textbf{93.09\%} \\
\bottomrule
\end{tabular}
\end{table}

% ════════════════════════════════════════════════════════════════
\section{Conclusion}

We presented a curriculum fine-tuning study transitioning a hybrid CNN-Transformer AMC model from a controlled, MIMO-expansion-cable-synchronized 915\,MHz base training to a real 4\,GHz free-space link across five distances, freezing the low-level CNN feature extractor and updating only the Transformer and task heads at each step. Matched-distance test accuracy remained in a narrow 91.8--93.7\% band across all five conditions, including the two partially-misaligned distances fine-tuned last in the curriculum, indicating that the adaptation strategy kept the model well-matched to the free-space link throughout. Confusion analysis identified two consistent, physically-motivated failure modes (QPSK/QAM4 constellation ambiguity and DSSS/BPSK chip-level similarity) present at comparable magnitude in every condition, suggesting they are properties of the class taxonomy rather than of the channel. Three ablations further support our design choices: a Transformer-free 1D-ResNet-only variant lost 2.28 points of validation accuracy relative to the full hybrid model, confirming the Transformer encoder's contribution; multi-task supervision outperformed single-task supervision at four of the five distances; and, while freezing the CNN stem during fine-tuning did not improve accuracy over updating the full network, it reduces updated parameters per fine-tuning step by $\sim$44\%, which we retain as a practical trade-off for resource-constrained onboard deployment. We reported honestly that our curriculum design confounds alignment with cumulative adaptation, and therefore make no claim to have isolated antenna-misalignment robustness as an independent effect. Future work includes a controlled, non-sequential fine-tuning design to separate distance, alignment, and adaptation-order effects, and extending the distance sweep beyond 75\,cm.

% ════════════════════════════════════════════════════════════════

\end{document}